\documentclass[conference]{IEEEtran}
\IEEEoverridecommandlockouts
\usepackage{cite}
\usepackage{amsmath,amssymb,amsfonts}
\usepackage{algorithmic}
\usepackage{graphicx}
\usepackage{caption}
\usepackage{nameref}
\usepackage{textcomp}
\usepackage{xcolor}
\usepackage{hyperref}
\def\BibTeX{{\rm B\kern-.05em{\sc i\kern-.025em b}\kern-.08em
    T\kern-.1667em\lower.7ex\hbox{E}\kern-.125emX}}

\begin{document}
\title{Droplet Simulations in Computer Graphics: Theories, Methods and Applications}

\author{\IEEEauthorblockN{Hossein Keshtkar*}
\IEEEauthorblockA{\textit{Mechanical and Aerospace Engineering} \\
\textit{Brunel University London}\\
London, UK \\
hossein.keshtkar@brunel.ac.uk}
~\\
\and
\IEEEauthorblockN{Nadine Aburumman}
\IEEEauthorblockA{\textit{Department of Computer Science} \\
\textit{Brunel University London}\\
London, UK \\
nadine.aburumman@brunel.ac.uk}
}

\maketitle
\begin{abstract}
Creating realistic droplet simulations and animations has long been a formidable challenge for researchers and developers due to the inherent complexity of fluid dynamics. Achieving lifelike droplet splash simulations while managing computational resources has often resulted in sacrifices compromising the realism of visualizations. Nevertheless, significant progress has been made in the past two decades, driven by advancements in particle-based methods such as Position-Based Dynamics (PBD) and Smoothed-Particle Hydrodynamics (SPH). These methods have enabled the simulation of droplet splash behaviour with increasing accuracy and reduced computational complexity. Integrating features like surface tensions, fluid incompressibility, and liquid-wall interactions has further enhanced the realism of the simulations. This paper provides an in-depth exploration of the theoretical foundations and methodologies employed in droplet simulations and how they have evolved over time. Accurate droplet interaction visualization holds immense potential across diverse applications, including gaming, animation, medical simulations, and engineering scenarios like 3D printing simulations.
\end{abstract}  
\section{Introduction}
\begin{figure*}[ht!]
      \includegraphics[width=\linewidth]{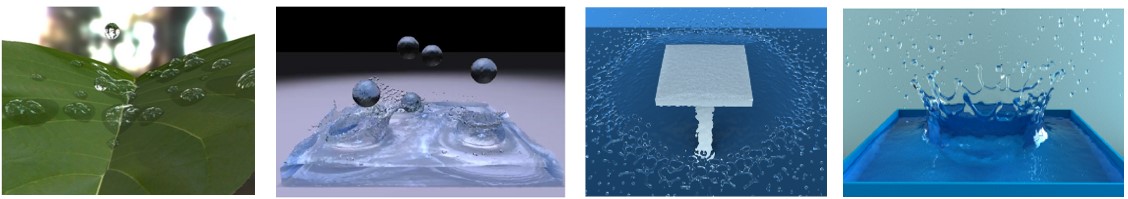}
    \caption{Due to the impact of solids and liquids, droplet simulations exhibit a range of microscopy phenomena, such as spreading, splashing, and crowning.}
    \label{fig:1}
\end{figure*}

Generating detailed visual simulations of droplet impact has been an active research topic within the computer science community for decades. Droplet impact visualization has a broad application ranging from animations to computer games, from engineering to medical simulations, it has always been important to visually replicate the actual physical behaviour of a droplet impact on a solid surface (see Fig.\ref{fig:1}). The ambition to visually simulate water drop behaviour dates back to decades ago when Dorsey et al. used a particle system to synthesize drops for large solid models, but it was assumed that each individual droplet’s deformation is too small to be noticeable \cite{Dorsey2005}. Later, Kaneda et al. developed a particle system to simulate water drops flowing on a flat surface \cite{Kaneda99,Kaneda1996}. Fournier et al. modelled flowing water drops using a mass-spring system with surface tension and volume conservation constraints. Although this method allowed various efficient simulations, it was unable to handle droplet separation and merging phenomena, especially when many droplets interact simultaneously on a domain \cite{Fournier1998}. Later, Yu et al. used a meatball concept to model static droplet shapes on flat surfaces successfully \cite{Yu1999} and Tong et al. used a similar concept to model water flows with a volume-preserving approach \cite{Tong2002} but none of these methods considered any surface tension effects on the moving interface, and thus neglected fluid dynamics within water droplets and surface liquid interfaces, resulting in an un-accurate representation of drop deformation and motion.

In order to capture more detailed interactions, engineers and researchers attempted to look into Finite Element Methods (FEM) and Volume-of-Fluids (VOF) methods as numerical techniques to model fluid droplet impact modelling. Although the implementations of the VOF go much further back \cite{Hirt1981}. To solve CFD (Computational Fluid Dynamic) problems and calculate accurate fluid behaviour, one needs to solve the Navier-Stokes equation [Eq.\ref{eq:1}]:

\begin{equation} \label{eq:1}
 \frac{dv}{dt} = \frac{-1}{\rho} \nabla P+\frac{1}{\rho}\nabla(\mu \nabla v)+g 
\end{equation}

The equation above is the most general version of the Navier-Stokes equation for fluid mechanics. However, using the FEM method required the Navier-Stokes equation to be discretized, and solving for the VOF method required many timesteps for coupled calculations. Even though these methods may have great precision and accuracy; due to the fact the VOF solves the continuity equation for every fluid and calculates the velocity field to track the fluid interfaces, and the FEM discretizes the computational domain and solves the Navier-Stokes equations for momentum and mass conservation and the energy equation for each discretized cell; they are very computationally demanding. These methods may be able to handle complex fluid-solid interactions with great accuracy, but they both require significant computational power, resources and time \cite{Bussmann2000}. Depending on the complexity of the simulation and the size of the computation domain, it may take several hours to simulate a droplet impact using FEM even with a very small-scale setup with basic geometries; whilst complex setups can easily take weeks to simulate.

\begin{figure}
  \includegraphics[width=\linewidth]{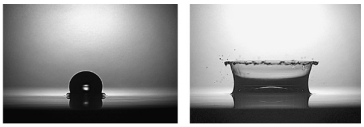}
   \includegraphics[width=\linewidth]{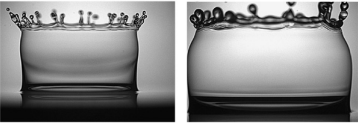}
    \includegraphics[width=\linewidth]{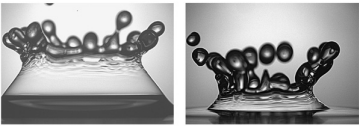}
  \caption{Droplet impact captured using high-speed photography techniques \cite{Sikalo2006}}
  \label{fig:2}
\end{figure}

The behaviour of impinging droplets is complicated, and depending on the circumstances, different characteristics may be observed. To name a few circumstances affecting the droplet impact behaviour, we can point to the size of the droplet, the angle of impact, the viscosity of the liquid, wetness or the dryness of the surface, flexibility or rigidity of the surface, the temperature of the surface, the velocity of impact, etc. which all can cause different behaviours from droplet impact dynamics \cite{Sikalo2006}. To capture all of these conditions in a detailed visual simulation is very challenging in terms of modelling, discretizing governing equations, computational resources and a critically important factor, computational time. Therefore, researchers investigated alternative approaches to allow detailed (see Fig \ref{fig:2}).

\begin{figure}
    \centering
      \includegraphics[width=\linewidth]{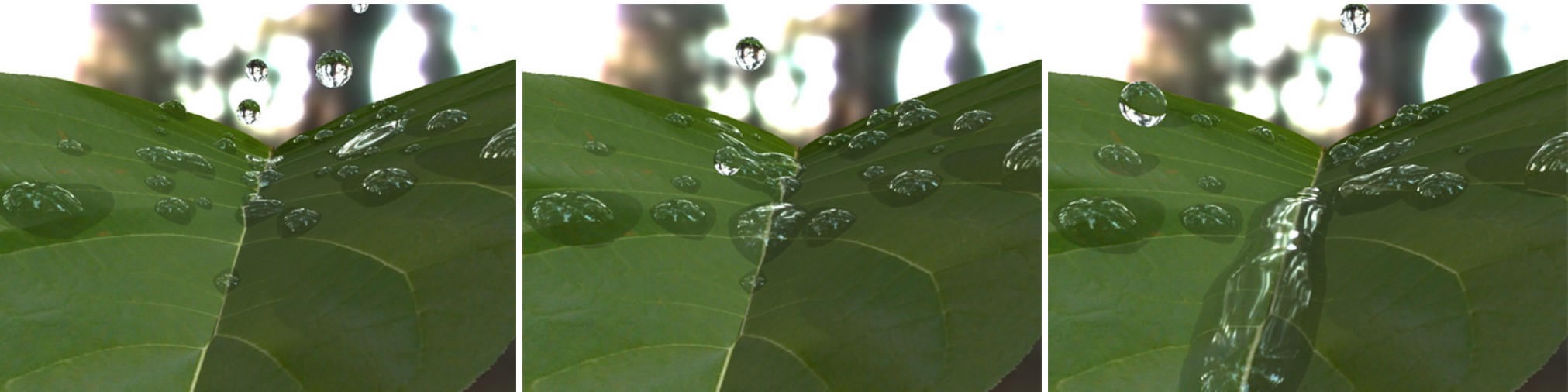}
    \caption{ Droplet simulation as shown by \cite{Wang2005}}
    \label{fig:3}
\end{figure}

Wang et al. adopted a different type of physically based method with a novel virtual surface approach that modified the level set distance field representing the fluid surface in order to maintain an appropriate contact angle \cite{Wang2005}. Using this method, Wang et al. showed easy handling of droplet breakup and coalescence, and since the liquid only occupies a small fraction of the whole solver domain, computational time and memory demand were reasonably low (see Fig \ref{fig:3}).

Another approach used for visualizing surface tension is the position-based dynamics (PBD) method, initially demonstrated by Macklin and Muller, which had the advantage of implementation easiness and computational performance \cite{Macklin2013}. The PBD framework relies on particle-based discretization with a density constraint on each particle to control the weighted average of the particle accumulation density in its vicinity. This gave the PBD method an advantage in simulating a broad spectrum of deformable bodies as demonstrated such as solids by Bender et al.\cite{Bender2014}, cloth by Kim et al. \cite{Kim2012}, rods by Umetani et al. \cite{Umetani2015}, sands by Macklin et al. \cite{Macklin2014}, various coupling effects by Frâncu and Moldoveanu \cite{Francu2017} and Rumann et al. \cite{AbuRumman2020}.

Despite the advantages of the PBD approach, the modelling of interfacial flow phenomena was largely unexplored, as this method uses an artificial pressure term to cohere the boundary particles inwards. Although this may be a simplifying factor in reducing computational demand, it may not be able to handle small-scale and thin fluid phenomena, such as films, bubbles or splash crowns, because naively increasing the artificial surface pressure or the number of iterations will not visually resemble such phenomena as they would appear in the real world.

To address this challenge, Xing et al. used a novel position-based dynamics (PBD) framework approach to capture detailed surface tension behaviour of liquid-film dynamics and droplet interactions \cite{Xing2022}. Using this method, Xing et al. showed the effectiveness of the enhanced model in simulating liquid tweezers, membrane filters and teapot effects with detailed surface tension visualization in only a few seconds of computational time. However, the drawback to their method was it strongly relied on very accurate surface particle detection. To such an extent that if an interior particle were to be accidentally identified as a surface particle, it would cause the whole system to act in a non-physical manner. Also, although Xing et al. contributions greatly added to the capabilities of the traditional PBD framework, it did not provide physically more accurate models to distinguish them from the initial principles; rather it broadened the application of the PBD framework to a wider range of simulation scenarios \cite{Xing2022}.

An alternative to the PBD method is the Smooth Particle Hydrodynamics method (SPH). Many implementations have been carried out on the SPH framework since it originally first emerged in 1977 as explained by Violeau and Rogers \cite{Violeau2016}. Modifications of the SPH framework usually resulted in improved versions of the model, each with added capacity to capture more realistic fluid interactions such as the work done by Fang et al. \cite{Fang2009}.

However, a drawback of the traditional SPH framework was the density underestimation in fluid-air or fluid-solid interfaces. In the traditional SPH framework, the surface particle densities are commonly incorrectly computed much lower than they should be, due to a lack of neighbouring particle forces. This in turn causes local negative pressure accumulation resulting in particle clustering on surfaces, also known as tensile instability in the SPH framework. Similar to the solution used for the PBD method, researchers such as Monaghan \cite{Monaghan2000}, and Macklin and Muller \cite{Macklin2013} tried applying an artificial pressure force on the surface boundary. However, this approach resulted in unrealistic fluid behaviour. There are also other approaches, such as density correction techniques proposed by Shepard \cite{Shepard1968} or the choice to simply not allow any negative density hotspots to occur to avoid tensile instability; but all these methods fail to show the fake physical behaviour of fluid interactions as they are missing critical components such as realistic pressure fields.

Different approaches were also tried such as curvature-based external forces on particles done by Muller et al. \cite{Muller2003}, or the modified SPH formulation by Clavet et al. \cite{Clavet2005}, and more recently Yu et al. proposed a forces-based method on surface mesh curvature \cite{Yu2012}. To resolve this challenge, researchers began looking deeper into surface tension characterization and modelling. Typically, surface tension modelling can generally be categorized into two subcategories: macroscopic and microscopic models. The Continuous Surface Force model (CSF), also known as the former, applies an asymmetric force to all fluid particles, but it may not guarantee momentum conservation. Meanwhile, the latter, the microscopic model, creates cohesion forces among particles to simulate real molecular attraction and repulsion forces for surface tension modelling.

Yang et al. managed to approximate capillary wave effects by converting surface tension energy changes measured from SPH methods into high-frequency density fluctuations and to counter any noise issue in wave simulation, a zero-pressure condition was applied on the free surface of the ISPH fluid domain \cite{Yang2016}.

In a novel approach, Akinci et al. demonstrated that if we applied a new inter-particle adhesion/cohesion force with a newly formulated surface tension force, in a fashion that these forces are applied to neighbouring fluid-fluid and fluid-boundary particle pairs symmetrically. This way, Akinci et al. prevented particle clustering at the free surface whilst satisfying the momentum conservation. Therefore, reproducing realistic fluid visualization such as water-crown formations and droplet interactions \cite{Akinci2013}. However, Wang et al. suggested the model proposed by Akinci et al. \cite{Wang2017} may have stability and unsatisfactory efficiency [30]. Instead, Wang et al. presented a combination of the Implicit Incompressible SPH (IISPH) method with the improved surface tension forces as presented by Akinci et al \cite{Akinci2013}; which claimed to create a good surface tension and adsorption effect, improving the computational efficiency and stability thus maintaining microscopic features.

Nair and Poschel proposed a different Incompressible SPH method to simulate 3D fluid-solid interactions, including surface wetting and surface tension, by taking into account the fluid capillary effect \cite{Nair2018}. Although Nair and Poschel showed impressive 3D crown formations resulting from the splashing of a droplet, but the ISPH method implemented was somewhat computationally demanding.
\section{Particle-based Methods}
\begin{figure}
  \includegraphics[width=\linewidth]{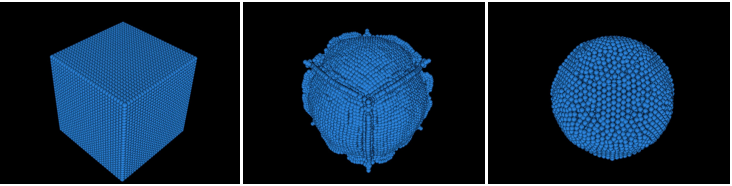}
  \caption{A demonstration of a particle-made droplet. \cite{Becker2007}}
  \label{fig:4}
\end{figure}

Particle-based approaches are a type of computational method used in fluid mechanics that studies the behaviour of fluids such as liquids and gases. In these approaches, the fluid is modelled as a collection of particles, each of which has a set of properties such as position, velocity, and density. The particles move in response to external forces such as pressure gradients and gravity (see Fig \ref{fig:4}), and their interactions with each other are governed by mathematical equations. This approach is known as the Lagrangian approach, named after Joseph-Louis Lagrange, a French mathematician and physicist who made significant contributions to the field of mechanics. In contrast to the Eulerian approach, which models the fluid as a continuous medium with properties that vary in space and time, the Lagrangian approach allows for a more detailed and dynamic analysis of fluid behaviour, making it a useful tool for simulating complex fluid systems.

\begin{figure}
  \includegraphics[width=\linewidth]{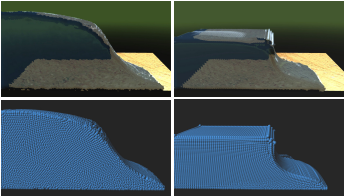}
  \caption{Combination of multiple particles to resemble liquid motion. \cite{Becker2007}}
  \label{fig:5}
\end{figure}

The Lagrangian method for particle tracking in fluid dynamics is a technique that is used to simulate the motion of individual particles in a fluid domain. This approach is based on tracking individual particles as they move through the fluid rather than tracking the fluid itself. The Lagrangian method involves calculating the position and motion of each particle individually based on the forces acting on it. These forces can include viscous, pressure forces, and other effects such as gravity or surface tension. The motion of the particle can be described using the equations of motion, which take into account the forces acting on the particle and its initial position and velocity.
To simulate the motion of particles in a domain using the Lagrangian method, one needs to know the velocity field acting on each individual particle, which can be calculated through numerical solutions. The position and velocity of each particle can then be calculated at each time step based on the velocity field and the equations of motion. The Lagrangian method is particularly useful for studying the behaviour of small particles in a fluid, such as pollutants spreading in the air, or a body of fluid splashing on a solid plane. These particles can experience complex motions due to the turbulent nature of the fluid, and their behaviour cannot be accurately described using traditional Eulerian methods that track the fluid as a whole.

A drawback of the Lagrangian method is that it can be computationally expensive since it requires tracking the motion of each individual particle. However, it is a powerful tool for understanding the behaviour of small particles in complex fluid flows, and it has many applications in environmental science, engineering, and other fields.

\subsection{Approximation of Smoothed Particle Hydrodynamics (SPH)}
\begin{figure}
    \centering
     \includegraphics[width=\linewidth]{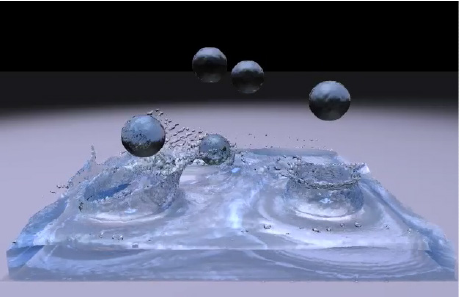}
  \caption{Exploiting GPU power to handle heavier SPH simulations as demonstrated by \cite{Harada2007}.}
  \label{fig:6}
\end{figure}

One of the common particle-based approaches using the Lagrangian method is Smooth Particle Hydrodynamics, also known as the SPH method. The SPH is a computational method used to simulate fluid behaviour in which the fluid is represented by a set of particles. The SPH method models the fluid as a collection of particles, with each particle having its own property, such as density, pressure, and velocity, which are calculated based on the properties of neighbouring particles. The idea of SPH is to follow individual particles in their motion. These particles can be viewed as material points carrying extensive quantities such as mass or volume, as well as intensive quantities such as velocity, pressure, turbulent kinetic energy (TKE) etc. Unlike traditional grid-based methods, the SPH method does not require a fixed grid mesh to represent the fluid, thus interpolation is based on particle position, by employing a smoothing function also known as Kernel; which is a function of the distance between the particle and neighbouring particles. This approach will allow the SPH method to easily handle complex geometries and free surfaces. Thus, the SPH method relies on several approximations to model the behaviour of the fluids using the particles. To increase the efficiency of SPH methods, some researchers worked on methods that could search for neighbouring particles on GPU's which could exploit significant computational GPU power, such as work done by \cite{Harada2007} (see Fig \ref{fig:6}).

As previously discussed, this method is based on fluid-particle discretization, and the number of discrete particles interacting within the fluid domain would be limited to the computational resources available. Although this may not be accurate in regions of high gradients such as shockwaves or boundary layers, this is an acceptable assumption for a broad range of simulations in engineering and computer graphic visualizations. To calculate the gradients of fluid properties, the SPH method uses a smoothing kernel function that gives weight to each particle based on its distance to the neighbouring particle \cite{Violeau2016}. This approximation assumes that the fluid properties vary smoothly within the domain space. The smoothing kernel function is commonly described as the cubic spline function by Lattanzio et al. \cite{Lattanzio1985}, which is calculated as [Eq.\ref{eq:2}]: 

\begin{equation} \label{eq:2}
  W(\overrightarrow{r},h)= ad
  \begin{cases} 
   \ \frac{2}{3} \ -q^2 + \frac{1}{2} q^3 &   1>q \geq 0 \\
   \frac{2}{3} - q^2 + \frac{1}{2}      &   2>q \geq 1 \\
  \end{cases}
\end{equation}
 
Where '$q$' is the ratio of the distance of two particles to the smoothing length. This kernel is used for almost all terms in the governing equations. However, for high-speed impact scenarios, special consideration needs to be taken when applying this function to pressure force calculations, as [Eq.\ref{eq:2}] will cause a reduction of pressure force when two particles move closer to each other. To avoid this problem, Johnson et al. developed a quadratic smoothing function to simulate high-speed impact problems \cite{Johnson1996}. This kernel is useful only for pressure calculations since its deviation will always increase as the particles move closer. This effect is critical for the repulsive force to be present, preventing particle clustering [Eq.\ref{eq:3}].

\begin{equation} \label{eq:3}
\begin{cases} 
  W(\overrightarrow{r},h)= a_d (\frac{3}{16}q^2 - \frac{3}{4}q +\frac{3}{4}) &   2>q\geq0 \\ 
\end{cases} 
\end{equation}

Muller et al. designed a different kernel structure to improve the SPH method \cite{Muller2003}. In this new employment, it was proposed for the smoothing function to have the shape of [Eq.\ref{eq:4}] with an exception for the pressure and viscosity forces, which needed careful consideration.

\begin{equation} \label{eq:4}
 W_poly6(\overrightarrow{r},h)=a_d
\begin{cases} 
  (1-q^2)^3 &   1>q\geq0\\ 
  0 & q>1
\end{cases} 
\end{equation}
 
And for the pressure and viscosity smoothing kernel, Muller et al. proposed [Eq.\ref{eq:5}] and [Eq.\ref{eq:6}] respectively.

 \begin{equation} \label{eq:5}
 W_spiky(\overrightarrow{r},h)=a_d
\begin{cases} 
  (1-q)^3 &   1>q\geq0\\ 
  0 & q>1
\end{cases} 
\end{equation}

\begin{equation} \label{eq:6}
 W_viscosity(\overrightarrow{r},h)=\frac{15}{2 \pi h^3}
\begin{cases} 
  \frac{-q^3}{2} + q^2 + q -1 &   1>q\geq0\\ 
  0 & q>1
\end{cases} 
\end{equation}
 
By adding the additional properties below, Mullet et al. made sure to prevent the kernel from creating negative Laplacian when two particles moved closer [Eq.\ref{eq:7},\ref{eq:8},\ref{eq:9}].

\begin{equation} \label{eq:7}
\nabla^2  W(\overrightarrow{r},h)= a_d (1-q)
\end{equation}

\begin{equation} \label{eq:8}
W(\lvert r \rvert =h,h)=0
\end{equation}

\begin{equation} \label{eq:9}
\nabla W(\lvert r \rvert =h,h)=0
\end{equation}
 
This way, the stability of the simulation was significantly increased, as reported by Fang et al. \cite{Fang2009}.
The basis of modelling fluid pressure and viscosity for a discretized particle domain is derived from the general Navier Stokes Equation as mentioned previously in [Eq.\ref{eq:1}].
Although different approaches have been proposed by researchers, with the intention of further increasing the effectiveness and applicability of the equation to reflect the physical behaviour of the particles. Fang et al. proposed a comprehensive formulation to discretize the Navier Stokes Equation \cite{Fang2009}.

\begin{equation} \label{eq:10}
\begin{split}
\frac{D\overrightarrow{\nu_i}}{Dt}=
\\
- \sum_{j=1}^{N} mj (\frac{Pi}{\rho ^2_i} + \frac{P_j}{\rho ^2_j}) \nabla_i
\\
 W_{ij} + \sum_{j=1}^{N} m_j(\frac{\tau_i}{\rho ^2_i} + \frac{\tau_j}{\rho ^2_j})
\\
\times \nabla_i W_{ij} +\overrightarrow{g} + \overrightarrow{f_s}
\end{split}
\end{equation}
 
Where the particle approximation for the viscous stress tensor, $\tau_i$, is

\begin{equation} \label{eq:11}
\begin{split}
\tau_i =  
\\
\sum_{j=1}^{N} \frac{m_j}{\rho_j} \mu_i \overrightarrow{\nu_{ij}} \nabla_i W_{ij} -
\\ 
\sum_{j=1}^{N} \frac{m_j}{\rho_j} \mu_i (\nabla_i  W_{ij})  \overrightarrow{\nu_{ij}}
\\
+ (\frac{2}{3}  \sum_{j=1}^{N} \frac{m_j}{\rho_j} \mu_i \overrightarrow{\nu_{ij}} \times \nabla_i W_{ij}) \delta_{ij}
\end{split}
\end{equation}

\begin{figure}
  \includegraphics[width=\linewidth]{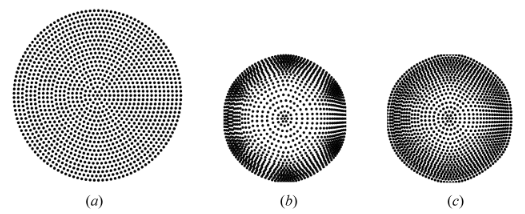}
  \caption{Simulation of a droplet shape. (a) initial distribution of SPH particles, (b) particle distribution after contracting using the equation of state, and (c) particle distribution after contration and implementing the pressure correction as described by [Eq. \ref{eq:10}] and [Eq. \ref{eq:11}]. \cite{Fang2009}}
  \label{fig:7}
\end{figure}
 
Whereas Wang et al. suggested a simpler approach with [Eq.\ref{eq:12} and Eq.\ref{eq:13}] \cite{Wang2017}:

\begin{equation} \label{eq:12}
f_i^P =  -\sum_{j}^{} m_j (\frac{P_i}{\rho_i^2} + \frac{P_j} {\rho_j^2}) \nabla W_{ij}
\end{equation}

\begin{equation} \label{eq:13}
f_i^\nu = \mu \sum_{j}^{} m_j \frac{\nu_{ij}}{\rho_j} \nabla^2 W_{ij}
\end{equation}
 
\subsection{The numerical integration methods that be used to solve SPH equations}
All the equations related to 
droplet impact dynamics is primarily characterized by the Reynolds number, and the Weber number which determine the relative significance of the inertial force to the viscous effect and the ratio of inertial to capillary forces, respectively, and are formulated by the following equations [Eq.\ref{eq:14},Eq.\ref{eq:15}] \cite{Karim2023}:

\begin{equation} \label{eq:14}
Re=\frac{\rho U D}{\mu}
\end{equation}

\begin{equation} \label{eq:15}
We=\frac{\rho U D^2}{\sigma}
\end{equation}
 
Where $\rho$ is the droplet density, $U$ is the droplet impact velocity, $D$ presents the droplet size (diameter), $\mu$ is the droplet viscosity and $\sigma$ is the surface tension of the fluid. Furthermore, the Ohnesorge number quantifies the balance between viscous force, inertia and the capillary force using [Eq.\ref{eq:16}]

\begin{equation} \label{eq:16}
Oh=\frac{\mu}{\sqrt{\rho U \sigma}} = \frac{\sqrt{We}}{Re}
\end{equation}
 
And the capillary number represents the comparative importance of the viscous force over interfacial tension using the equation below [Eq.\ref{eq:17}].

\begin{equation} \label{eq:17}
Ca=\frac{\mu U}{\sigma}
\end{equation}
 
To simulate visual characteristics of droplet impact, we would have to follow the physical formulations governing the fluid droplet behaviour. Furthermore, to solve the domain within the Lagrangian framework, we have discretized our bulk domain into numerous individual particles. Thus, we need to discretize the governing equations accordingly, to reflect the behaviour of each particle. As mentioned in subsection 2.1.1, to formulate the SPH method, the Navier-Stokes Equations [Eq.\ref{eq:1}] is used, which is the most general form of fluid motion. In order to discretize this equation, making it suitable for application within the Lagrangian framework over numerous individual particles, Fang et al. (see Fig \ref{fig:7}) wrote the Navier-Stokes equation as [Eq.\ref{eq:10}] \cite{Fang2009}. 
 
Tartakovsky et al. rewrote [Eq.\ref{eq:11}] and exchanged the viscosity terms with their respective coefficients to construct [Eq.\ref{eq:18}] \cite{Tartakovsky2016}.

\begin{equation} \label{eq:18}
\begin{split}
\frac{dU}{dt} = -\sum_{b}^{} m_b \ (\frac{P_a}{\rho_a^2} 
+ \frac{P_b} {\rho_b^2}) \nabla_a W_{ab}
\\
+ \sum_{j}^{} m_b \ (\frac{2\mu}{(\rho_a \rho_b)}  \frac{r_{ab} . \nabla W_{ab}} {r^2_{ab} +
\epsilon^2})U_{ab} + f^{int}_a + f^b_a
\end{split}
\end{equation}

Where $m$ is the particle mass, $\rho$ is the density, $p$ is the pressure at a particle identified by the subscript $a$ or its neighbour $b$. In this equation, $R_{ab}$ is the displacement vector between particle $a$ and $b$. The function $W$ is the smoothing kernel function. The pairwise forces $f^{int}_a$ represents the interfacial forces and $f^b_a$ denotes the body forces per unity of mass acting on particle $a$, which usually is only the gravity force. Within the above formulations, commonly an extensive version of the viscous force (which is the second term on the right) is used [Eq.\ref{eq:19}] as initially suggested by Morris et al. \cite{Morris1997}:

\begin{equation} \label{eq:19}
\nabla \cdot \left( \frac{\mu}{\rho} \nabla u \right)_a =
\sum_b m_b \left( \frac{\mu_a + \mu_b}{\rho_a \rho_b} \frac{\mathbf{r}_{ab} \cdot
\\
 \nabla_a W_{ab}}{r_{ab}^2 + \epsilon^2} \right) U_{ab}
\end{equation}
 
In which $U_{ab}$ is the relative velocity vector between particle $a$ and $b$, and $\epsilon$ is a very small positive number which is used to avoid division by zero in rare cases of particles overlapping each other \cite{Nair2018}.

\subsection{Incompressible Smoothed Particle Hydrodynamics (ISPH)}
Whilst using SPH formulation in the form of [Eq.\ref{eq:10}] may seem suitable, this formulation is derived from the most general form of the fluid behaviour, thus taking into account fluid compressibility as it is based on the ideal gas equation. Although compressibility and fluctuations in density is a natural phenomena for gaseous particles, but for cases of liquid or droplet simulation this characteristic can cause undesired and visually unreal oscillations within the fluid particles. This is very natural considering the nature of fluids being almost incompressible. However, enforcing incompressibility for particle methods is a challenging problem. Using traditional SPH methods, the computed density of particles at the fluid-air interface is calculated lower than its real value, which is caused by a lack of neighbour particles. This results in the generation of negative pressure and causes particle clustering. In addition, the simulation of microscopic characteristics takes a large amount of calculation and has the problem of time step restriction and numerical instability. 
To solve this issue, Cumins and Rudman proposed a projection approach for the Lagrangian methods that are also used in Eulerian approaches to tackle this issue \cite{Cumins1999}. 
Becker and Teschner also presented a weakly compressible form of the SPH method based on the Tait equation using pairwise forces based on cohesion. To solve the compressibility issue, they formulated the density function as [Eq.\ref{eq:20}] \cite{Becker2007} (see Fig \ref{fig:5})

\begin{equation} \label{eq:20}
    \rho_a= \sum_b m_b W_{ab}
\end{equation}
 
In which $W_{ab}=W(x_a-x_b)$ .

In [Eq.\ref{eq:20}] since the mass is carried by the particle itself, the mass conservation is constantly applied. However, due to the fact that surface particles in free surface scenarios may have fewer neighbouring particles, this equation could cause incorrect lower-density predictions near the surface. Several approaches have been proposed to solve this issue. For instance, Sigalotti et al. used an adaptive kernel length $h$ to enforce constant density at the surface in order to reconstruct small-scale effects at the surface \cite{Sigalotti2006}. Also, there are various forms relating pressure and density for fluid states. To apply the incompressibility factor in these states, most commonly, the Poisson equations [Eq.\ref{eq:21}] have to be solved as demonstrated by Premoze et al. \cite{Premoze2003}.

\begin{equation} \label{eq:21}
    \nabla^2 P= \rho \frac{\nabla \nu}{\Delta t}
\end{equation}
 
Premoze et al. suggested a realized incompressibility for the Moving Particles Semi-implicit method. This method involves projecting a velocity estimate onto a pressure-corrected, divergence-free subspace, which requires solving the Poisson equation. Despite its ability to handle large time steps, solving the Poisson equation using the conjugate gradient solver is time-intensive for larger systems. Premoze et al. noted 3 minutes of computational time for 100k particles, while compressible SPH implementations can process similar complexities in 5-10 s per time step.
In contrast to Premoze et al. Becker and Teschner proposed using the Tait equation, which enforces very low-density variations and is computationally efficient. The Tait equation has the form of [Eq.\ref{eq:22}] which is in the form of relative pressure, and to calculate the absolute pressure one needs to simply add the constant atmospheric pressure into the equation.

\begin{equation} \label{eq:22}
    P=B ((\frac{\rho}{\rho_0})^\lambda - 1)
\end{equation}
 
In the equation above, Becker and Teschner suggested $\lambda=0.7$ and $B$ is the pressure constant governing the relative density fluctuations. This method, which allowed small user-set density fluctuations, was in contrast to the previous strictly incompressible SPH methods, which were time-consuming to solve the Poisson equation. The improved method proposed by Becker et al. could also visualize effects such as the splashing and breaking of waves naturally; although this method was particularly appropriate for single-phased free surface flows \cite{Becker2007}. 

\subsubsection{Free surface simulation with SPH}
The free surface flow involves the presence of a free interface between two different fluids, such as water and air. In the SPH methodology, the free surface is represented as a discontinuity in density and pressure, where the boundaries of one fluid end and the next begin. To consider this boundary change, SPH uses surface tension force which is derived from the curvature of the liquid and is modelled as an attractive force, which we will cover in the next subsection.
Traditionally, the SPH method does not require a boundary condition at the free surface since all particles move according to their Lagrangian velocity. Also, the surface position and form is only resolved up to the particle sizes. However, calculating the pressure for the particles on the free surface using traditional methods may cause ambiguous calculations or incorrect estimations due to the limited number of neighbouring particles. Several approaches have been proposed to tackle this problem. For example, Nair and Tomar suggested for ISPH, in which the pressure is being computed from a linear system, the pressure of the free surface particles must be forced to be zero by applying a Dirichlet boundary condition \cite{Nair2014}. However, this approach would require extensive tracking of the surface particles, which may cause fluctuations in the results.
If a dynamic free surface condition was implemented with the WCSPH, it would manage to automatically set the surface pressure to zero thanks to [Eq.\ref{eq:23}], which is governed by this condition.
\begin{equation} \label{eq:23}
    P=f(\frac{\rho}{\rho_0} , P_B , C_0)
\end{equation}
 
where $\rho_0$ is a reference density, $P_B$ is background pressure and $C_0$ a numerical speed of sound \cite{Violeau2016}.
Also for weakly compressible SPH methods, Colagrossi et al. showed that different smoothing operators would be required to account for the incomplete kernel support if the free surface boundary conditions are not explicitly imposed at the surface \cite{Colagrossi2009}. Becker and Teschner also worked with weakly compressible SPH methods for free surface flows \cite{Becker2007}. They looked at the phenomena from a microscopic point of view, in contrast to the common macroscopic view. This way, they developed a surface tension model that we will further discuss in the next section, which generated an attractive force between the surface molecules which reflected surface tension, modelling and visualizing the free surface flow even with high curvatures. They suggested this attractive force to be smoothed by a kernel according to [Eq.\ref{eq:24}]

\begin{equation} \label{eq:24}
    \frac{dV_a}{dt}=-\frac{\kappa}{m_b} \sum_{b} m_b W (x_a - x_b)
\end{equation}
   
Xing et al. proposed a new scheme for identifying surface particles in weakly compressible SPH methods using changes in density ratios and three auxiliary functions. Xing et al \cite{Xing2012}. suggested this new scheme would considerably improve the identification of particles on the free surface without sacrificing computational duration.
Another approach for defining the free surface is by employing dummy particles adjacent to the free surface, acting with a repulsive force holding the free surface intact. These dummy particles prevent the penetration of surface particles through the boundary wall. In spite of the simplicity of this approach, the initial placement of several layers of dummy particles can be difficult, especially for non-geometric computational domains as explained by Shobeyri \cite{Shobeyri2018}. Shobeyri then proposed two different and more efficient approaches. In the first strategy, Shobeyri implemented the contribution of dummy particles in the numerical computations based on their prior known positions. In the second strategy, Shobeyri excluded all the dummy particles entirely and used the wall particle method to model wall conditions, which were described to have improved efficiency and accuracy in the dam-break flow simulation \cite{Shobeyri2018}.

\subsubsection{Surface tension simulation with SPH}
The surface tension plays an important role in modelling fluid and droplet interaction as the surface is the contact point of the two fluids, commonly water and air, and the characteristic behaviour of this will have significant visual effects on the overall display.  As previously mentioned, using the SPH method within the Lagrangian framework involves modelling numerous individual particles which interact with each other to represent the whole liquid domain. The position and value of each particle are calculated in relation to its neighbouring particles. However, on the free surface due to the lack of neighbouring particles, traditional SPH methods may seem inaccurate as they will cause particle clustering and surface oscillations. This is because such forces on the fluid surface will appear as inter-molecular forces in contrast to the scalar fields, like pressure, which could be evaluated much more easily on a macroscopic aspect. Typically surface tension in liquids arises as a result of inter-molecular cohesion. However, the simple SPH methodology treats the fluid particles on a macroscopic level with a, thus the cohesion forces in between particles are not considered, and thus the surface tension visualized is not what we anticipate to see in reality. This is because only the very few neighbouring particles will attract each other, causing local clustering effects on the surface. Also, it has been shown by Muller et al. that enforcing curvature minimization terms on the surface alone will result in more severe particle clustering, causing simulated droplets to break into smaller droplets unnecessarily \cite{Muller2003}. To overcome this challenge, researchers looked into various approaches to model the surface tension to keep the physical behaviour as close to reality as possible. 
For example, Tartakovsky and Meakin were one of the first pioneers to employ a molecular cohesion force to generate surface tension using SPH methodology \cite{Tartakovsky2005}. To do so, they made modifications to the cosine function to generate an attraction force on particles further apart from each other whilst creating a repulsion force for closer particles. However, observations showed this approach also resulted in particle clustering to some extent.

Later, Becker and Teschner replaced the cosine function with a SPH kernel function in an attempt to improve the surface behaviour. However, this method also suffered from clustering as it lacked repulsive terms \cite{Becker2007}. These attempts used the displacement vector between the neighbouring particles by considering a support radius of $h$. Therefore, if the distance between the particles was reduced under this support value, the repulsive forces would vanish for very close particles; which is physically in-correct, prompting the clustering effect. 
To solve this issue, Akinci et al. proposed a novel approach to handle these challenging scenarios \cite{Akinci2013}. They introduced a two-way adhesion attraction force on fluid-fluid and fluid-boundary pairs in a symmetrical way. Akinci et al used an alternative cohesion force formulated as [Eq.\ref{eq:25}]
\begin{equation} \label{eq:25}
F^{cohesion}_{i-j}=-\gamma \ m_i \ m_j \ C \ (\lvert x_i-x_j \rvert \frac{x_i-x_j}{\lvert x_i - x_j \rvert})
\end{equation}
 
Where $i$ and $j$ are the neighbouring particles, $m$ is the particle mass and $x$ is the position of the respective particle, $\gamma$ is the surface tension coefficient and $C$ is a spline function developed by Akinci et al for 3D SPH simulations as \cite{Akinci2013}:

\begin{equation} \label{eq:26}
C(r)= \frac{32}{\pi h^9}
    \begin{cases} 
(h-r)^3 r^3 &   2r>h \quad  and \quad  r \leq h\\
2(h-r)^3r^3 - \frac{h^6}{64} & r>0 \quad and \quad 2r\leq h\\
0 & otherwise,
\end{cases} 
\end{equation}
 
\begin{figure}
    \centering
    \includegraphics[width=\linewidth]{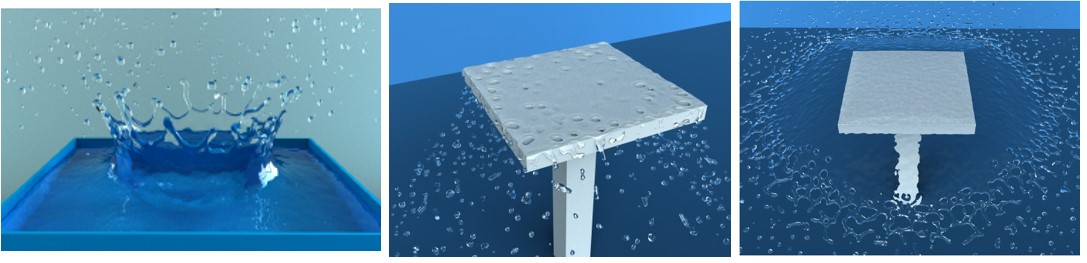}
    \caption{Droplet interaction simulated by \cite{Akinci2013}, where their simulation could capture  phenomena such as water crown formation.}
    \label{fig:8}
\end{figure}

Using this method, the inter-particle attraction forces exist until two particles reach a rest distance of $\frac{h}{2}$, at which point the attraction force smoothly vanishes as the particles draw closer, and the repulsion force starts generating and building stronger. This way, the particles auto-adjust to balance out the forces. This was a novel approach as both the attraction and repulsion forces behaved like a Gaussian effect to avoid clustering. Also, the newly introduced repulsion forces did not disappear for very close neighbouring particles, which significantly prevents clustering for underestimated pressure regions. Furthermore, a notable improvement of the Akinci et al. method was that repulsion term stopped increasing beyond a certain point after two particles moved very close, which helped in avoiding over-stiff forces and improved the stability issues.
 The benefit of this method is it excluded other techniques such as using ghost air particles to make-up for the lack of surface neighbouring particles, or complex surface tracking techniques. Akinci et al. concluded that with their approach of combining the adhesion and cohesion forces in a paired symmetrical manner, they could handle real-life scenarios such as water crown formation or various fluid-solid interactions plausible. Wang et al. built on the model proposed by Akinci et al. and presented a surface tension-adhesion method on implicit ISPH method to solve the problem of not considering adhesion in these cases \cite{Wang2017}. To do so, Wang et al. considered molecular cohesion and surface minimization techniques and combined them with adhesion to show the microscopic characteristics of the surface. Wang et al. reported their method had higher efficiency and better stability than the original Akinci et al. model, and it could show micro characteristics such as minimizing the fluid surface in a more effective way (see Fig \ref{fig:8}). Although this method was impressive on liquid-air simulations, but for multi-phased fluids where the surface tension is the dominant force acting on the particles, this approach may not be as effective.\

\section{Grid-based Methods}
There are two distinct techniques for discretizing a physics problem to enable simulation: Lagrangian and Eulerian methods. Whilst in the previous section, we discussed the Lagrangian method in great detail, in this section, we will introduce and briefly explain the Eulerian method. This method involves discretizing the domain space in which the material flows, in contrast to the Lagrangian method, which involves discretizing the material itself. Thus, within the Eulerian context, the material moves through a fixed grid meshes in space. All Eulerian methods are based on a meshed domain, such as finite volumes or finite elements method. The simulation consists of separate volume elements that cover the entire domain. While these volume elements don’t necessarily form a regular grid, it is often the preferred choice for implementation and performance reasons to form a uniform grid in the space. Since the discretization of space is independent of the material, the algorithms can easily handle significant deformations in the solver domain; consequently, these class of simulation algorithms are well suited for fluid simulation and remain prevalent in computer graphics \cite{Muller2007}.

One challenge of Eulerian simulations is that the discretization of the simulation domain remains constant regardless of the material's shape, making it difficult to track interfaces and handle moving boundaries. Additionally, numerical dissipation can result in mass loss.
The most prominent algorithms in this category are mesh-based finite element methods, widely employed in continuum mechanics. The numerical accuracy of these methods depends significantly on the aspect ratio of the mesh elements used to cover the simulated material, typically tetrahedra. If the material undergoes substantial deformations, the mesh quality deteriorates, necessitating remeshing.

\subsection{Advection}
To simulate fluid interactions, including droplet impact, we need to solve the principle Navier-Stokes equation [Eq. \ref{eq:1}]. Solving the advection equation simply involves solving the velocity and acceleration terms, see [Eq. \ref{eq:27}].

\begin{equation} \label{eq:27}
\frac{Du}{Dt} = 0
\end{equation}

To encapsulate [Eq. \ref{eq:27}] in a numerical routine, one can write it in the form of [Eq. \ref{eq:28}]

\begin{equation} \label{eq:28}
u^{n+1}= advect (\overrightarrow{q}, \Delta \ t , q^n)
\end{equation}

Which effectively will discretize the velocity field $\overrightarrow{u}$ over a timestep of $\Delta  t $, and the current field quantity $q^n$ returns an approximation to the result of advecting $q$ through the velocity field over that duration of time. [CITE FLUID SIMULATION BOOK]

One obvious solution to solving [Eq. \ref{eq:27}] for a $\Delta t$ time interval would be to simply write the Navier-Stokes equation in a PDE form [Eq. \ref{eq:29}]. This is a different layout to the former principle equation introduced in [Eq. \ref{eq:1}]. This new partial derivative layout is acquired as it is well adapted to the discretized grid-based Eulerian method.

\begin{equation} \label{eq:29}
\frac{\partial u}{\partial t} + u . \nabla \ u + \nabla \ P = 0
\end{equation}

By using the PDE layout introduced above, we can easily replace the derivatives with finite differences to solve the fixed mesh domain. For example, a common method would be the forward Euler method which uses accurate central difference for the spatial derivative against a fixed time interval [Eq. ]

\begin{equation} \label{eq:30}
\frac{q^{n+1}_i - q^n_i}{\Delta t} + u^n_i \frac{q^{n}_{i+1} - q^n_{i-1}}{2 \Delta x} = 0
\end{equation}

Which can be arranged into an explicit formula to solve for the new values of q [Eq. \ref{eq:31}]

\begin{equation} \label{eq:31}
q^{n+1}_i = q^n_i - \Delta t \ u^n_i \ \frac{q^n_{i+1} - q^n_{i-1}}{2 \Delta x}
\end{equation}

The above forward Euler method was an example of solving the discretized advection field, whereas there exist other numerical models, to solve this equation such as the upwind method, or the QUICK method and etc. Explaining these models in depth is outside the scope of this paper; however, we encourage readers to follow the published lecture notes from Iaccarino on this topic \cite{Iaccarino2004}.

\subsection{Solving the Pressure Term}
Unlike gases, which can be easily compressed and their volume reduced, liquids maintain their volume almost unchanged even when subjected to significant pressure. This property stems from the close proximity of liquid particles and their strong intermolecular forces. When pressure is exerted on a liquid, these intermolecular forces counteract the compressing force, preventing the particles from coming closer together. As a result, the liquid retains its volume and only undergoes minimal changes in density. 
To solve the pressure terms in the Navier-Stokes equation for droplet impact, several approaches can be employed depending on the specific scenario and desired level of accuracy. Here, we'll discuss two common methods: the potential flow approximation and the numerical methods.

1) Potential Flow Approximation: \\
The potential flow approximation assumes that the flow is irrotational, meaning that the vorticity (rotational motion) of the fluid is negligible. This assumption simplifies the Navier-Stokes equation by considering only the pressure and velocity terms. In the context of droplet impact, this approximation is often valid for low-viscosity fluids and low Reynolds numbers.

By assuming potential flow, the Laplace equation can be used to solve for the velocity potential, which is related to the velocity field of the fluid. Once the velocity potential is known, the pressure distribution can be obtained by applying Bernoulli's equation [Eq. \ref{eq:32} ] or the boundary conditions of the specific droplet impact problem.

\begin{equation} \label{eq:32}
P_1 + \frac{1}{2} \rho v^2_1 + \rho g h_1 = P_2 + \frac{1}{2} \rho v^2_2 + \rho g h_2
\end{equation}

2) Numerical Methods: \\
Numerical methods provide a more general and versatile approach to solving the Navier-Stokes equation for droplet impact. These methods involve discretizing the fluid domain into a computational grid and approximating the partial differential equations with numerical schemes.

Finite difference, finite volume, and finite element methods are commonly employed numerical techniques. These methods divide the domain into discrete cells or elements and solve the Navier-Stokes equation iteratively at each point in the grid. The pressure terms are determined by solving a pressure Poisson equation, which is derived from the discretized momentum equations.
Numerical methods allow for more realistic simulations of droplet impact by considering the full range of fluid behaviors, including vorticity, turbulence, and complex geometries. They can handle high Reynolds numbers, non-Newtonian fluids, and other complex flow characteristics. Examples of this approach can be found in \cite{Dunkel2014}  \cite{Moghtadernejad2020} \cite{Josserand2016}.

To capture more realistic liquid behaviour, the Navier-Stokes solver has to be equipped with tools to reflect the incompressibility behaviour of liquids. To apply incompressibility to the solver, Xu and Ren \cite{Xu2022} applied the “projection” step to the existing Navier-Stokes equation, which included acquiring the pressure field by solving the Poisson equation [Eq. \ref{eq:33}].

\begin{equation} \label{eq:33}
\nabla\ . \nabla\ P = \frac{1}{\nabla\ t} \nabla\ . \ u
\end{equation}

The velocity field correction is applied with the gradient of the pressure as [Eq. \ref{eq:34} ]

\begin{equation} \label{eq:34}
u_{t+1} = u_t - \Delta \ t \nabla P_t 
\end{equation}

\begin{figure}
  \includegraphics[width=\linewidth]{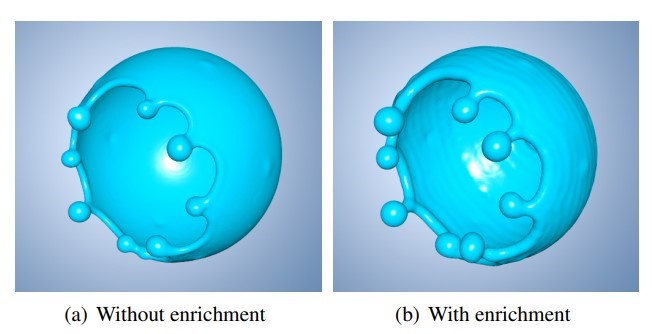}
  \caption{A bursting bubble. (a) using SPH without enrichment, (b) using enrichment as explained by Yang et al. \cite{Yang2016}.}
  \label{fig:9}
\end{figure}

Solving Equation \ref{eq:27} is essentially equivalent to resolving a substantial and sparse linear system, which becomes exponentially larger as the input size grows. Commonly, iterative methods such as PCG and Jacobi are employed to tackle this mathematical problem. However, as the input size expands, both the memory and computation costs increase. Additionally, the number of iterations required for convergence grows significantly. Nevertheless, achieving high-resolution simulations of real-world fluid dynamics remains a persistent objective in practical applications. To overcome the challenges presented using this approach attempts such as the ones done by Tompson et al. and Yang et al. have been done to replace traditional pressure solutions with deep learning approaches by introducing convolutional neural networks to explore the distribution and correspondence rules of fluid divergence and pressure fields. While the convolution operation significantly reduces the size of trainable parameters and computation costs in comparison to a fully connected layer, the computation and time demands escalate considerably as the input fluid size expands, particularly when simulating the fluid field in a three-dimensional domain (see Fig \ref{fig:9}). This not only hampers the performance of the solution but also restricts the ability to design a more efficient network structure \cite{Tompson2017, Yang2016}.

Other researchers such as Xu and Ren used the existing approach as proposed by \cite{Tompson2017} and \cite{Yang2016}, and redesigned the network structure to make it flexible in order to achieve comparable outcome accuracy with significantly lower computational cost. 
The researchers applied convolutional multiplications in the fluid field and gradually extracted features in the coarse grid to solve for a pressure field. To align the pressure field solution with the original resolution, an additional connection layer was designed to fuse the coarse grid features with the original fluid quantities. This alignment strategy within the network provided the fluid simulation with the same level of detail while significantly reducing the computation cost. By performing the most convolutional manipulations on the coarse grid, a large-scale computation cost was effectively reduced \cite{Xu2022}. 

\subsection{Incompressible Fluid}

To ensure incompressibility, [Eq. \ref{eq:29}] needs to be coupled with $ \nabla. \ \overrightarrow{u} =0$ \ . On the domain grid, we will approximate this condition with the finite differences and require that the divergence estimated for every grid cell be zero for $\overrightarrow{u}^{n+1} $. The divergence in the three-dimensional form can be expressed as [Eq. \ref{eq:35}] written in partial derivatives suitable for grid discretization.

\begin{equation} \label{eq:35}
\nabla . \overrightarrow{u} = \frac{\partial u}{\partial x} + \frac{\partial v}{\partial y} + \frac{\partial w}{\partial z}
\end{equation}

The above equation could be easily discretized using common methods to cover the entire 3-dimensional grid domain. For example, if we assume the central differencing scheme, we would arrive at [Eq. \ref{eq:36}].

\begin{equation} \label{eq:36}
\begin{split}
(\nabla . \overrightarrow{u})_{i,j,k} = \frac{u_{i+1/2,j,k} - u_{i-1/2,j,k}}{\Delta x} 
\\
+ \frac{v_{i,j+1/2,k} - v_{i,j-1/2,k}}{\Delta y} + \frac{w_{i,j,k+1/2} - w_{i,j,k-1/2}}{\Delta z}
\end{split}
\end{equation}

Applying fluid mechanics principles to droplet interaction visualization allows capturing very detailed effects over a range of flexible conditions which can be independently imposed on the simulation, such as adding external forces, advecting velocity fields, computing diffusion and applying incompressible projection operations. However, a common problem is that an Eulerian grid's first-order integration and finite resolution would cause numerical artefacts in fluid simulations. To tackle this issue, Stam developed stable Eulerian methods with many additional efforts to enhance fluid flow reality \cite{Stam99}.  Dupont et al. and Selle et al. have proposed methods to estimate the underlying numerical error within this method and correct it, reducing the numerical deficiencies of the grid structure \cite{Dupont2003, Selle2008}. This way, the visual results of the Navier-Stokes solver were improved by refining the underlying numerical methods. 

The graphics community typically divides the process of solving the Navier-Stokes equations into three steps: advection, force application, and projection. During the advection step, the fluid's properties are transported based on its velocity. This means that the fluid's attributes, such as density or temperature, are carried along with the flow. In the force application step, external forces are introduced to the fluid for a short period of time. These forces can include gravity, wind, or any other influences acting on the fluid. The purpose of this step is to simulate the effects of external forces on the fluid's behaviour \cite{Xu2022}. 

\subsection{Boundary Conditions}

The boundary conditions affecting droplet impact on solid surfaces depend on the specific physics being considered and the numerical method employed. Different researchers adopted different tactics to capture liquid droplet interaction on impact. In this section, we will introduce the most common boundary conditions governing the state of a droplet at the moment of impact.

1) No-slip Condition: \\
The no-slip boundary condition is a fundamental requirement when modelling droplet impact on a solid surface. It assumes that the fluid velocity at the solid surface is equal to the surface velocity, meaning that the droplet's velocity relative to the surface is zero. This condition arises from the strong viscous forces that exist at the fluid-solid interface. This condition is commonly applied to capture the interaction between the droplet and the solid surface. The no-slip condition is usually always considered in order to visualize liquid-solid interactions such as a splash effect. By enforcing the no-slip condition, the simulation captures the effect of adhesion between the droplet and the solid surface, ensuring that the droplet comes to rest at the impact point and experiences the necessary momentum transfer during impact. The no-slip boundary condition is essential for accurately predicting droplet spreading, splashing, and other impact characteristics as proven by many research such as \cite{Riboux2017}.

2) Wall Temperature: \\
The wall temperature boundary condition plays a significant role in modelling droplet impact, particularly when considering heat transfer during the process. By specifying the temperature at the solid surface, the simulation can accurately capture the thermal behaviour and its influence on the droplet impact dynamics. The wall temperature affects several aspects of droplet impact, including the evaporation or condensation of the droplet, the spreading behaviour, and the overall heat transfer rates. Depending on the relative temperature difference between the droplet and the wall, the wall temperature boundary condition can result in various outcomes, such as droplet cooling, vaporization, or even phase change. Properly accounting for the wall temperature allows for an accurate representation of the heat transfer mechanisms during droplet impact and provides insights into thermal phenomena associated with applications like spray cooling, combustion, or thermal management systems.
The wall temperature boundary condition also affects the wetting behaviour of the droplet during impact. The temperature gradient between the droplet and the solid surface can influence the spreading or receding of the droplet on the wall. Higher wall temperatures, for example, can promote droplet evaporation and reduce contact time, leading to faster spreading or even complete splashing. On the other hand, lower wall temperatures can result in enhanced wetting, longer contact times, and slower spreading \cite{Schmidt2022}. Therefore, proper consideration of the wall temperature boundary condition is crucial for accurately predicting the wetting behaviour and understanding the interplay between thermal effects and droplet impact dynamics. By accounting for the wall temperature, simulations can provide insights into a wide range of applications, including spray cooling, combustion systems, and heat transfer processes involving droplet impact.

\begin{figure}
  \includegraphics[width=\linewidth]{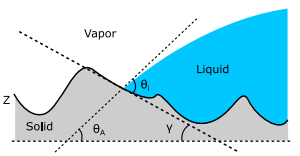}
  \caption{Schematic view of surface roughness on droplet point of contact \cite{Gelissen2020}.}
  \label{fig:10}
\end{figure}

3) Surface Roughness: \\
The surface roughness boundary condition has a notable impact on droplet impact simulations, as it influences the interaction between the droplet and the solid surface. The presence of surface roughness alters the local flow field and affects the spreading behaviour, contact area, and splashing tendency of the impacting droplet. The roughness elements on the solid surface introduce additional drag forces, causing deviations from the idealized smooth surface behaviour (see Fig \ref{fig:10}). Depending on the roughness characteristics, such as height, distribution, and spacing of roughness elements, the droplet may experience enhanced spreading due to increased contact area or encounter obstacles that hinder its spreading and promote splashing. Accurately capturing the surface roughness boundary condition is crucial for predicting the droplet impact behaviour in scenarios where the solid surface exhibits non-smooth or textured features, which could also affect secondary droplets in cases such as icing accretion \cite{Wang2021}.

The surface roughness boundary condition also affects the wetting behaviour of the droplet during impact. Rough surfaces can alter the contact angle and wetting characteristics of the droplet. The presence of microstructures or irregularities on the solid surface can promote partial wetting, leading to the formation of pinned contact lines or contact angle hysteresis. This behaviour significantly influences the spreading dynamics, contact time, and receding behaviour of the droplet. By incorporating an appropriate surface roughness boundary condition, the simulation can accurately capture the complex interplay between droplet impact, surface roughness, and wetting behaviour, enabling insights into phenomena observed in practical applications, including additional implied drag forces as researched by \cite{Chung2021}

4) Wetting Condition: \\
The wetting condition also plays a critical role in modelling droplet impact, as it governs the behaviour of the droplet at the fluid-solid interface. The wetting condition is typically described in terms of the contact angle, which represents the angle between the droplet's liquid-vapour interface and the solid surface. The wetting behaviour strongly influences the spreading, recoiling, and splashing tendencies of the droplet during impact. Usually, the surface wetting conditions are closely related to wall temperature conditions if they both coexist, as has been shown by \cite{Das2020}. By specifying the contact angle as part of the wetting condition, the simulation can capture the dynamic wetting behaviour, including partial wetting, complete wetting, or even contact angle hysteresis \cite{Ahmed2014}, \cite{Gao2006}. The wetting condition affects the contact area, spreading rate, and final shape of the droplet. 

The wetting condition boundary condition also takes into account the presence of contact line effects. The contact line is the three-phase interface where the liquid-vapour interface meets the solid surface. Proper treatment of the contact line is vital for accurately capturing the wetting behaviour during droplet impact. The wetting condition boundary condition needs to account for the dynamics of the contact line, which can exhibit complex behaviour. By appropriately defining the wetting condition boundary condition, the simulation can visually represent the interfacial dynamics at the contact line, leading to a more realistic depiction of the droplet impact process and enabling insights into the wetting phenomena observed in various practical applications. Usually, the surface wetting conditions are closely related to wall temperature conditions if they both coexist, as has been shown by \cite{Das2020} making it a crucial factor in understanding the impact dynamics and visual effects a droplet will display.

5) Contact Line Treatment: \\
The Contact Line Treatment boundary condition is of utmost importance when modelling droplet impact, as it addresses the behaviour of the contact line where the liquid-vapour interface meets the solid surface. The contact line treatment is crucial because the contact line dynamics significantly influence the spreading behaviour, receding, and overall wetting characteristics of the droplet during impact. Accurate modelling of the contact line is challenging due to the complex interplay of capillary forces, viscous forces, and intermolecular interactions at the microscopic scale. Various contact line models and boundary conditions have been developed to capture the dynamic behaviour of the contact line \cite{Chamakos2016} and \cite{Griebel2014}, such as using the dynamic contact angle \cite{Kumar2022} or incorporating slip conditions at the contact line. By properly implementing the Contact Line Treatment boundary condition, the simulation can accurately represent the intricate behaviour of the contact line, leading to more realistic predictions of the spreading dynamics, splashing phenomena, and wetting behaviour observed in droplet impact scenarios.

\begin{figure}
  \includegraphics[width=\linewidth]{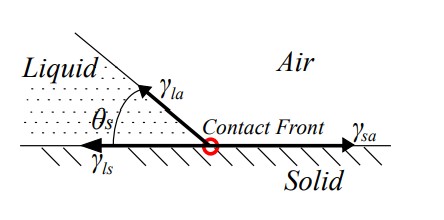}
  \caption{Contact angle of a liquid droplet on a solid surface. \cite{Wang2005}}
  \label{fig:11}
\end{figure}

The Contact Line Treatment boundary condition also influences the energy balance and heat transfer aspects during droplet impact (see Fig \ref{fig:11}). As the droplet spreads on the solid surface, the evaporation or condensation processes occur at the contact line. The heat transfer rates and the local temperature distribution at the contact line are influenced by the contact line dynamics and the wetting behaviour. Properly capturing the contact line treatment is crucial for accurately modelling the heat transfer mechanisms, such as the heat flux at the contact line, the evaporative cooling effect, or the generation of vapour near the contact line. By incorporating appropriate boundary conditions to address the contact line treatment, the simulation can provide insights into the intricate interplay between the contact line dynamics, wetting behaviour, and heat transfer phenomena during droplet impact, enabling a more comprehensive understanding of the overall impact process.

It's important to note that the choice of boundary conditions may vary depending on the specific numerical method, software, and physical phenomena being considered in the simulation. To ensure physical realism, it is crucial to carefully select and implement appropriate boundary conditions based on the specific physics, numerical methods, and solver approaches to obtain meaningful and insightful results from droplet impact simulations.

\section{Applications of Droplet Simulations}

Droplet simulations play a crucial role in computer graphics, enabling the realistic depiction of various natural phenomena involving liquids, such as rain, water splashes, and even complex fluid dynamics. These simulations are essential in creating visually captivating and believable scenes in movies, video games, virtual reality experiences, and other computer-generated imagery.

One of the primary applications of droplet simulations is in creating realistic splash effects. Water droplets falling can be simulated using physics-based algorithms, considering factors like gravity, velocity, and collision with surfaces. This level of realism adds depth and immersion to scenes, enhancing the overall visual quality. Also, these effects could be used to simulate effects such as raindrops
and the splash it makes on impact \cite{Feng2005,Rumman2015}. Whether it's a serene drizzle or a torrential downpour, droplet simulations help replicate the behaviour of raindrops as they interact with the environment and various objects in the scene.

Moreover, droplet simulations are invaluable when it comes to generating water splashes and fluid interactions. Whether it's a character diving into a pool, a stone skipping across a pond, or a powerful ocean wave crashing against rocks, droplet simulations accurately model the fluid dynamics involved. This level of precision in depicting water behaviour not only makes scenes more visually appealing but also aids in storytelling, as the interactions with water can be critical elements in many narratives.

Another exciting application lies in the creation of interactive virtual environments. With droplet simulations, users can interact with virtual water bodies in real time, allowing for immersive experiences in virtual reality (VR) and gaming. The ability to generate realistic water surfaces, droplet effects, and fluid interactions enhances the sense of presence and engagement, creating captivating and convincing digital worlds. Using realistic simulations in VR is not limited to the video gaming experience, as research has shown huge potential and promise in integrating such methods for teaching and learning applications \cite{Cai2014}. 

Further, wide applications have been identified in using droplet splash emulation for medical and engineering simulations \cite{Subramaniam2021}. The importance of visualizing droplet behaviour in medicine and engineering range from tracking disease spread via liquid droplets (such as Covid spread during oxygen therapy) to additive manufacturing and simulating 3-D printing applications \cite{Sun2023}, or numerical investigations of water droplet impact on aircraft wing structure \cite{Manigandan2019}.

In summary, droplet simulations are a powerful tool in computer graphics, enabling the creation of visually stunning and immersive content. By accurately modelling the behaviour of liquids like rain, water splashes, and fluid dynamics, these simulations contribute to the realism and believability of virtual environments and add an extra layer of authenticity to movies, video games, VR experiences, and other forms of digital media. 
\section{Conclusion}
In this paper, we have reviewed the advancements in droplet simulation techniques within the computer graphics community that have significantly contributed to the ability to generate realistic and computationally efficient visualizations of complex fluid behaviours. Particle-based methods such as Smoothed Particle Hydrodynamics (SPH) and Position-Based Dynamics (PBD) have demonstrated versatility in capturing intricate droplet interactions, including coalescence, separation, and surface tension effects while addressing computational constraints. These methods, along with hybrid approaches and grid-based techniques, enable the modelling of diverse scenarios such as droplet splashes, fluid-solid interactions, and free-surface flows. The continuous improvement of simulation frameworks, including the development of enhanced surface particle detection methods and advanced incompressibility formulations, has broadened the applicability of these models. Furthermore, the integration of machine learning for pressure field calculations highlights an emerging direction in reducing computational costs while maintaining simulation fidelity.

Applications of droplet simulation span across entertainment, engineering, and medical fields, enhancing realism in gaming and films, aiding in additive manufacturing processes, and facilitating studies on droplet behavior in medical contexts. The increasing adoption of data-driven techniques and hybrid modelling approaches promises to address limitations such as computational expense and resolution challenges in simulating small-scale or multi-phase phenomena.

Future simulations consider the convergence of physical modelling and data-driven methodologies could offers exciting opportunities for achieving unprecedented levels of detail and efficiency in droplet simulation. As these technologies evolve, their potential to enrich visual simulation and provide practical insights across disciplines will undoubtedly expand.
\bibliographystyle{IEEEtran}
\bibliography{STARDropletSimulation}         
\end{document}